\documentclass[conference]{IEEEtran}
\IEEEoverridecommandlockouts
\usepackage{cite}
\usepackage{amsmath,amssymb,amsfonts}
\usepackage{graphicx}
\usepackage{textcomp}
\usepackage{xcolor}
\usepackage{cite}
\usepackage{amsmath,amssymb,amsfonts}
\usepackage{graphicx}
\usepackage{textcomp}
\usepackage{listings}
\usepackage{xcolor}
\usepackage{amssymb}
\usepackage{amsmath}
\usepackage{caption}
\usepackage{subcaption}
\usepackage{algorithm}
\usepackage{algpseudocode}
\usepackage{enumerate}
\usepackage{booktabs}
\usepackage{tabularx}
\usepackage{array}
\usepackage{booktabs}
\usepackage{tcolorbox}
\usepackage{pifont}
\usepackage{fvextra}
\usepackage{lstlinebgrd}
\usepackage[numbers]{natbib}
\usepackage{fancyhdr}

\definecolor{dkgreen}{rgb}{0,0.6,0}
\definecolor{gray}{rgb}{0.5,0.5,0.5}
\definecolor{mauve}{rgb}{0.58,0,0.82}
\definecolor{mygray}{gray}{0.9}

\definecolor{lightblue}{HTML}{b2dfdb}
\newcommand*{\mtcolor}{\color{lightblue}}

\def\BibTeX{{\rm B\kern-.05em{\sc i\kern-.025em b}\kern-.08em
    T\kern-.1667em\lower.7ex\hbox{E}\kern-.125emX}}
\begin{document}

\title{An Empirical Study on Benchmarks of Artificial Software Vulnerabilities}


\author{\IEEEauthorblockN{Sijia Geng\IEEEauthorrefmark{1}\quad
Yuekang Li\IEEEauthorrefmark{2}\quad Yunlan Du\IEEEauthorrefmark{1}\quad
Jun Xu\IEEEauthorrefmark{3}\quad Yang Liu\IEEEauthorrefmark{2}\quad Bing Mao\IEEEauthorrefmark{1}}
\IEEEauthorblockA{\IEEEauthorrefmark{1}Nanjing University} 
\IEEEauthorblockA{\IEEEauthorrefmark{2}Nanyang Technological University}
\IEEEauthorblockA{\IEEEauthorrefmark{3}Stevens Institute of Technology}
}

\maketitle

\thispagestyle{fancy} 
\lhead{} 
\chead{} 
\rhead{} 
\lfoot{} 
\cfoot{\thepage} 
\rfoot{} 
\renewcommand{\headrulewidth}{0pt} 
\renewcommand{\footrulewidth}{0pt} 
\pagestyle{fancy}

\begin{abstract}
Recently, various techniques (e.g., fuzzing) have been developed for vulnerability detection. To evaluate those techniques, the community has been developing benchmarks of artificial vulnerabilities because of a shortage of ground-truth.
However, people have concerns that such vulnerabilities cannot represent reality and may lead to unreliable and misleading results.
Unfortunately, there lacks research on handling such concerns.
 
In this work, to understand how close these benchmarks mirror reality, we perform an empirical study on three artificial vulnerability benchmarks - LAVA-M, Rode0day and CGC (2669 bugs) and various real-world memory-corruption vulnerabilities (80 CVEs).
Furthermore, we propose a model to depict the properties of memory-corruption vulnerabilities.
Following this model, we conduct intensive experiments and data analyses. 
Our analytic results reveal that while artificial benchmarks attempt to approach the real world, they still significantly differ from reality.
Based on the findings, we propose a set of strategies to improve the quality of artificial benchmarks.

\end{abstract}


\section{Introduction}

Software vulnerability has long been a fundamental threat against 
computer security. To alleviate this situation, tremendous research
has been conducted on developing techniques to find software 
vulnerabilities. In the last five years, over 60 papers about 
vulnerability detection have been published on top conferences 
of cyber security and software engineering~\cite{klees2018evaluating}.

To comprehensively measure the utility and understand the limitations 
of techniques of vulnerability detection, the community has been developing 
artificial benchmarks of software vulnerabilities. People compose 
such benchmarks by either injecting vulnerabilities into 
existing programs~\cite{dolan2016lava,Rode0dayWeb} or crafting 
artificial vulnerable programs~\cite{CGCWeb}. Compared with benchmarks 
consisting of real-world programs, artificial benchmarks have two 
advantages. On the one hand, artificial benchmarks can provide 
ground truth of vulnerabilities, which is essential 
for the evaluation of vulnerability coverage. On the other hand, 
artificial benchmarks can synthesize diversities around the 
vulnerabilities. 
Such diversities can evaluate the 
generality of vulnerability detection techniques.

Despite the above advantages, while some bug finding tools  perform well on artificial benchmarks, they may not perform  well in real-world programs. For example, Wang et al.\cite{WangXLC2020TGF} find that although Angora (a new mutation-based fuzzer) can detect almost all the bugs that inserted by LAVA while AFL performs worse, for some real-world vulnerabilities, AFL can discover more vulnerabilities (e.g., CVE-2017-6966, CVE-2018-11416, CVE-2017-13741) than Angora within 24 hours across 8 runs.
So researchers are hesitating 
with adopting artificial vulnerability benchmarks, concerning that 
these benchmarks insufficiently represent reality and thus,  
evaluation using such benchmarks is unreliable or even 
misleading. Unfortunately, there lacks research on handling such 
concerns in the literature. 

In this work, we aim to understand the validity of artificial 
benchmarks. Our methodology is to compare the mainstream benchmarks of 
artificial vulnerabilities with real-world 
vulnerabilities and focus on the following questions and seek answers through
our study:\\
{\bf Q1.} \emph{How do artificial vulnerabilities differ from real-world vulnerabilities? Can current artificial vulnerabilities sufficiently mirror 
the reality?}\\ 
{\bf Q2.} \emph{What can we find by modifying the artificial benchmarks according to what we have observed from Q1?} \\
{\bf Q3.} \emph{What improvements can we make towards more realistic 
artificial vulnerabilities benchmarks?}

To answer the above questions, we develop a general model that captures the essential properties of vulnerabilities.
This model depicts the properties of memory corruption vulnerabilities, together with how the properties influence the evaluation on the vulnerability detection tools, as we will detail in Section~\ref{sec:model}. 
Following this model, we carry out our experiments and summarize answers to the above three questions. 
In short, our study reveals that while artificial benchmarks attempt to approach the real world, they still differ from the reality.
For \textbf{Q1}, we found that the artificial benchmarks created by injecting bugs do not capture the key properties of real-world vulnerabilities in terms of bug requirements and vulnerability types.
For \textbf{Q2}, we found that the artificial bugs cannot mirror the reality well because they may not fairly reflect the program state coverage of the vulnerability detection tool.
For \textbf{Q3}, we propose several strategies to make artificial benchmarks more realistic based on our analyses and experiments.

In summary, our major contributions are four-fold as follows.
\begin{itemize}
    \item First, to the best of our knowledge, we perform the first in-depth empirical study on the analysis of benchmarks of artificial vulnerabilities with manual verifications and confirmations. 
    Our large-scale analysis covers 2669 artificial bugs and 80 real-world vulnerabilities, which is the most comprehensive comparison between artificial vulnerabilities and real-world ones.
    \item Second, we develop a general model to describe essential software vulnerabilities, together with how the bug requirements influence the evaluation on the vulnerability detection tools. 
    \item Third, our results provide quantitative evidence on the differences between artificial and real-world vulnerabilities from some properties (e.g., requirements to trigger a vulnerability).
    Also, we identify how the properties influence the evaluation of vulnerability detection techniques.
    \item Fourth, we modify the benchmarks to make it more realistic according to what we have observed on control flow and data flow. 
    We also propose new improvements toward making artificial benchmarks more realistic.
\end{itemize}

\section{Background and Motivations}

\label{section:Background}

In this section, we start by introducing the background of security vulnerabilities and benchmarks we study.
We then proceed to describe our research goals.

\subsection{Memory Corruption Vulnerability}

A memory corruption vulnerability is a security vulnerability that allows attackers to manipulate in-memory content to crash a program or obtain unauthorized access to a system.
Memory corruption vulnerabilities such as "Buffer-Overflow", "Null-Pointer-Dereference" and "Use-After-Free", have been ranked among the most dangerous software errors~\cite{CWEdanger}.

\subsection{Benchmarks of Artificial Software Vulnerabilities}
\label{sec:dataset_bg}

\begin{figure}[hb]
	\centering
	\includegraphics[width=0.5\textwidth]{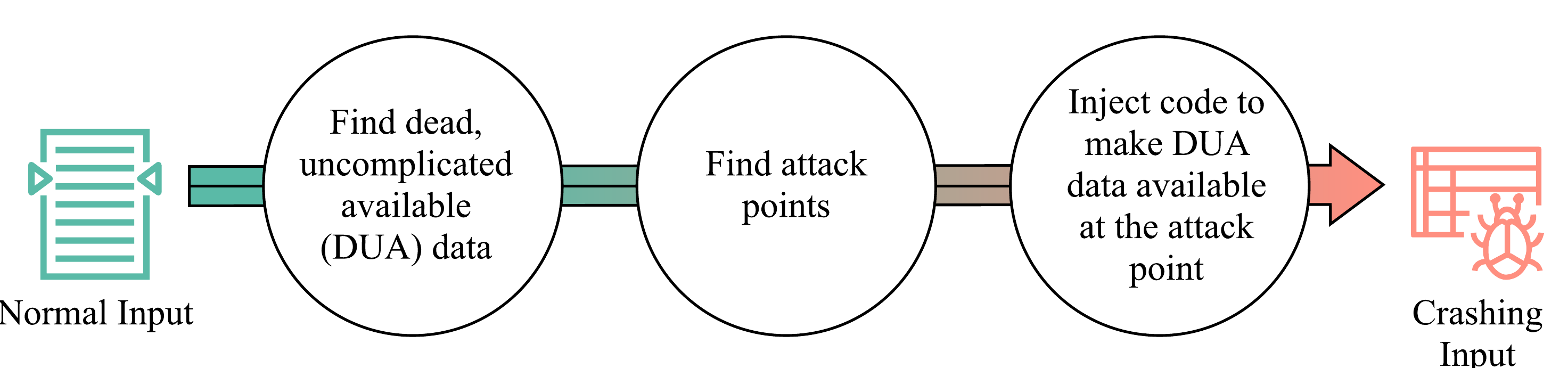}
	\caption{An overview of how LAVA plants bugs.}
	\label{fig:LavaOverview}   
\end{figure}

In spite of decades of research in bug detection tools, there is a surprising dearth of ground-truth corpora that can be used to evaluate the efficacy of such tools.
The lack of ground-truth datasets means that it is difficult to perform large-scale studies of bug discovery.

To comprehensively measure the utility and understand the limitations of vulnerability detection techniques, the community has been developing artificial benchmarks of vulnerabilities.  
In recent years, three widely-used artificial vulnerability databases have sought to address the need for ground-truth benchmark and there are two categories of them. 
\textbf{LAVA-M} and \textbf{Rode0day} inject large numbers of bugs into program source code automatically and include similar bugs structurally.
And \textbf{DARPA Cyber Grand Challenge} (CGC) contains a series of fairly small programs with known bugs written by security experts.
In this part, we will briefly provide background information on the artificial vulnerabilities before introducing our study.

LAVA is a prevalent ground-truth corpus generator, which can efficiently and automatically add memory corruption bugs into C/C++ programs on demand. 
LAVA currently concentrates on injecting Buffer-Overflow bugs and produces corresponding proof-of-concept inputs as their triggers~\cite{dolan2016lava}.

Figure~\ref{fig:LavaOverview} presents an overview of LAVA. 
When LAVA obtains the source code and series of input files of the program, it finds unused portions of the input by running the program with dynamic taint analysis~\cite{newsome2005dynamic} on each specific input.
This data parts of the inputs bytes are \emph{dead} (i.e., it does not influence control flow), \emph{uncomplicated} (not modified), and \emph{available}, namely DUAs. 
Since DUAs are often a direct copy of input bytes and can be set to any chosen value without sending the program along a different path, LAVA intends to regard them as candidate triggers for memory corruption. 
At the DUA site, code is inserted to copy the DUA's value into a global variable via a call to {\tt lava\_set}.
Then LAVA finds potential attack points (ATPs), which is a program instruction involving a memory read or write whose pointer can be modified and must occur temporally after a DUA along the program execution.
LAVA introduces a dataflow relationship between DUA and ATP along the trace by modifying the source code as follows and at the attack point, {\tt lava\_get} retrieves the value, compares it to the trigger value (magic number). 
If the condition is met, then a buffer-overflow will occur.

\textbf{LAVA-M}, generated by LAVA, is one of the datasets in our study. It is a benchmark of artificial software vulnerabilities that injected more than one bug at a time into the source code and used widely to evaluate vulnerability detection tools~\cite{stephens2016driller, li2017steelix, rawat2017vuzzer, chen2018angora}. 

\textbf{Rode0day}~\cite{Rode0dayWeb} is a recurring bug finding contest, which is also a benchmark to evaluate vulnerability discovery techniques. 
In the Rode0day contest, the successful detection of memory error vulnerabilities is demonstrated in the form of a simple crash.
Evolved from LAVA, Rode0day uses automated bug insertion to generate new error assemblies in the form of standard 32-bit Linux ELF files, to help evaluate the performance of vulnerability detection tools.

\textbf{CGC}~\cite{CGCWeb}, a competition among autonomous vulnerability analysis systems, is a widely adopted benchmark in recent studies~\cite{chen2018angora,li2017steelix,rawat2017vuzzer,zhao2019send}.
This competition resulted in a collection of small programs with known vulnerabilities and triggering inputs.
Each challenge of CGC contains one or more bugs, which are deliberately devised by security experts to evaluate vulnerability detection tools.

\subsection{Motivations}

While existing works focus on proposing new benchmarks that can be used to evaluate the efficacy of vulnerability detection tools~\cite{roy2018bug, dolan2016lava, Rode0dayWeb, CGCWeb, pewny2016evilcoder}, there is a lack of systematic understanding of whether the artificial vulnerabilities can represent the real-world vulnerabilities.
Although other works suggest that some widely-used benchmarks, like LAVA-M, Rode0day and CGC are different from real-world bugs in several ways~\cite{roy2018bug, hu2018chaff, pewny2016evilcoder}, they did not provide in-depth analysis of the benchmarks.
Moreover, there is also an absence of analysis on how the different properties of the benchmark vulnerabilities can reflect the different features of the vulnerability detection tools.
Also, analyzing the differences between artificial and real-world vulnerabilities can help to create more realistic benchmarks for tool evaluation.
Thus, we propose a study to provide a first in-depth understanding of similarities and differences between artificial and real-world vulnerabilities while exploring solutions to make artificial bugs more realistic.

\section{Model Overview}
\label{sec:model}

\begin{figure*}[h]
	\centering
	\includegraphics[width=\textwidth]{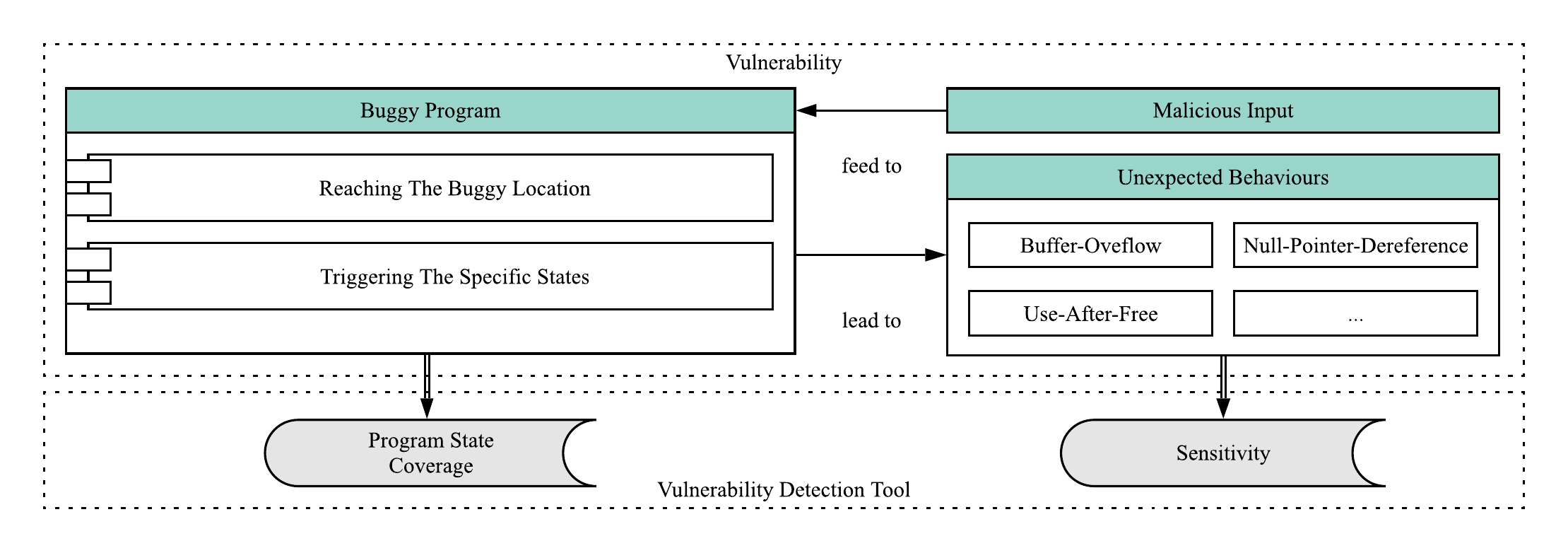}
	\caption{The model for memory corruption vulnerability and vulnerability detection tool.}
	\label{fig:model}   
\end{figure*}

In this paper, we focus on the memory corruption vulnerability, which is the majority of software security defects~\cite{ms_mc}.
Instead of discussing the vulnerabilities case by case, we propose a \emph{general model} to describe the memory corruption vulnerabilities by summarizing the \emph{requirements for triggering the vulnerability}.
Besides the model for vulnerabilities, we also propose a model to represent the different \emph{features of the vulnerability detection tool} and how they are \emph{affected by} the components of the vulnerability model.
This is because the goal of the benchmarks is to evaluate the vulnerability detection tools.
So we further extend the vulnerability model by adding the relations with the properties of the vulnerability detection tool.
Figure~\ref{fig:model} depicts our model, in which the cyan boxes represent the main components of a bug and the gray boxes are the properties of the vulnerability detection tools.
This model provides a systematic basis for our experiments and analyses.
Here we discuss the model in detail.

From Figure~\ref{fig:model}, we can see that a typical memory corruption vulnerability happens like this: first, the target program is fed with a malicious input; then, the program must contain bugs/vulnerabilities which can be triggered by the malicious input; last, after the vulnerability is triggered successfully, the program will behave unexpectedly.

Triggering the unexpected behaviour is the coaction of the malicious input and the vulnerability inside the program.
In our model, we focus on the requirements for the program to become vulnerable.
Such requirements are called \textbf{bug requirements} in our model.
The bug requirements can be summarized as follows.
To trigger the vulnerability, we need to feed a specific input to the buggy program.
This input can be in different forms, such as a file, a series of user actions or a stream of data, but it should satisfy the following two conditions.
\ding{182} By executing this input, the control flow must \textbf{reach the buggy location} in the program.
Nevertheless, the reachability problem is hindering the detection of most vulnerabilities.
For example, the control flow needs to go through a series of complex conditions and the data flow needs to go through a series of data transformations.
\ding{183} In further, this input may have to lead the program's execution to the location with \textbf{a specific program state}.
In other words, only reaching the buggy location is \emph{not} enough.
For example, for a Use-After-Free vulnerability, the program may have reached the location of the buggy ``\emph{use}" without passing through the corresponding ``\emph{free}".
In this case, the vulnerability is not triggered.
To trigger the Use-After-Free vulnerability, the program must reach the ``\emph{use}" from the path where the ``\emph{free}" has been called.
To summarize, a vulnerability is triggered when the buggy program is fed with an input that can help to reach the buggy location with a specific program state.


After triggering the vulnerability, the buggy program will \textbf{behave unexpectedly} in different ways.
The program may run into Buffer-Overflow, Null-Pointer-Dereference, Use-After-Free, etc. and harm the software and system.

Since the goal of the artificial benchmarks is to evaluate different
properties of vulnerability detection tools, our model also shows how the different components of vulnerabilities are connected with the properties of tools.
Both reaching the buggy location and triggering specific states require the tool to use some strategies to solve the control-flow and data-flow constraints to have a larger \emph{program state coverage}.
Also, some unexpected behaviours of the buggy programs require the tool to be sensitive enough to detect.
For example, if a tool can only detect one type of vulnerability and other types cannot detect, then the tool is insensitive.
So the vulnerability types can evaluate the \emph{sensitivity} of tools.

To sum up, the high-level model contains the above properties mentioned and forms the basis of our study. 
Following the general model, we design our information extraction methodology in Section~\ref{section:DATASETS AND METHODOLOGY} and discuss the details of three research questions in Section~\ref{sec:analysis_results}.

\section{Datasets and Methodology}

\label{section:DATASETS AND METHODOLOGY}

In this section, we describe the dataset for our experiments and high-level information extraction methodology.

\subsection{Datasets}
\label{sec:datasets}

\begin{table}[htbp]
\centering
\caption{Dataset overview.}
\label{tab:dataset-overview}
\begin{tabular}{c|c|c|c|c|c}
\toprule
\textbf{Dataset}& \textbf{Real World} & \textbf{LAVA-M} & \textbf{Rode0day} & \textbf{CGC} & \textbf{Total} \\ \hline
\textbf{Bugs Number} & 80                  & 587          & 1963            & 119              & 2749\\ 
\bottomrule
\end{tabular}
\end{table}



\textbf{Artificial Benchmarks.}
We choose three widely-used artificial benchmarks, which are LAVA-M, Rode0day and CGC. 
LAVA-M and Rode0day inject bugs into programs automatically.
We collect 587 bugs in LAVA-M and 1963 bugs in Rode0day.
And LAVA-M insert bugs into 4 open-source programs ({\tt base64}, {\tt md5sum}, {\tt uniq} and {\tt who}) while Rode0day insert bugs into 9 programs.
For CGC, the specialist-written benchmark, we select 93 challenges with 119 memory corruption bugs (Table~\ref{tab:dataset-overview}).

\textbf{Real-world programs.}
We analyze 80 representative CVEs~\cite{CVEWeb} which are from 40 different open-source software products (eg. {\tt ngnix}, {\tt openjpeg}, {\tt perl}, {\tt php}).
The reason that we focus on memory corruption vulnerabilities is because of their high severity and significant real-world impact.
We list the CVE-IDs of 80 real-word vulnerabilities at the website~\cite{StudyOnBenchmarks}.
We obtained the types from the CVE entry's description or its references.

\subsection{Creation of the dataset}

Note that, the distribution of different category of memory corruption
vulnerabilities nearly follows the statistics we collect from
the real world, which is shown in Table~\ref{tab:CGC-Realworld-Bug-Type}. 
To make the real-world vulnerabilities more extensive and the analytic results more general, we divide the vulnerabilities into 5 times bins(3 - year period) whose over-time trend of our dataset relatively consistent with that of the entire CVE dataset~\cite{CVEdatabase}.
To sum up, the creation of the dataset is consist with the 
aforementioned aspects and the real-world circumstances, 
hence the dataset is of rationality and representativeness.
The 80 real-world vulnerabilities with CVE-IDs are shown at the website.


\begin{figure}[htbp]
\centering
\begin{minipage}{0.435\linewidth}
\includegraphics[width=\linewidth,height=1.2in]{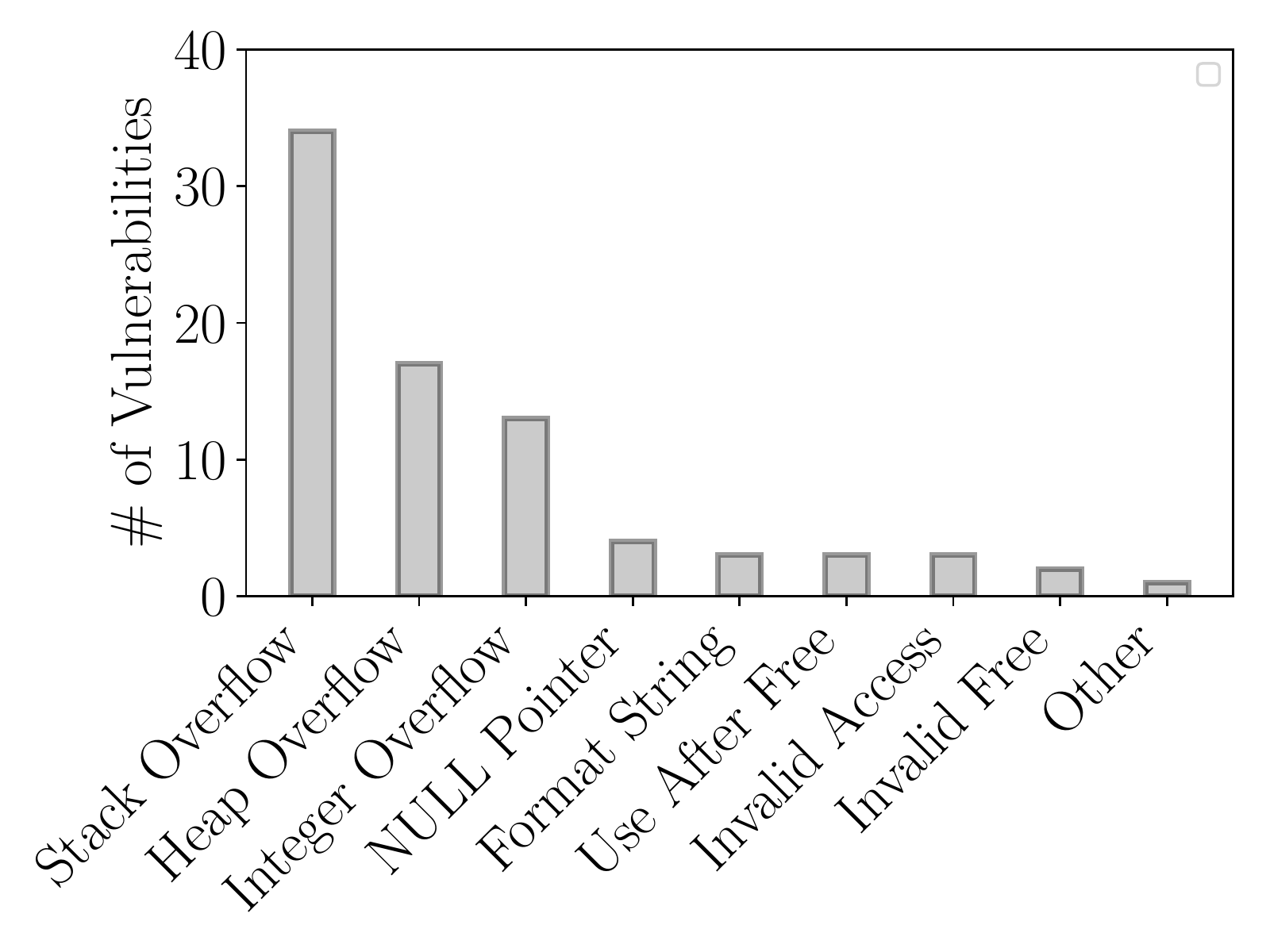}
\vspace{-0.2in}
\caption{Real-world vulnerability type of our dataset.}
\label{fig:Vulnerability-type}
\vspace{-0.03in}
\end{minipage}
\hfill
\begin{minipage}{0.45\linewidth}
 \includegraphics[width=\linewidth,height=1in]{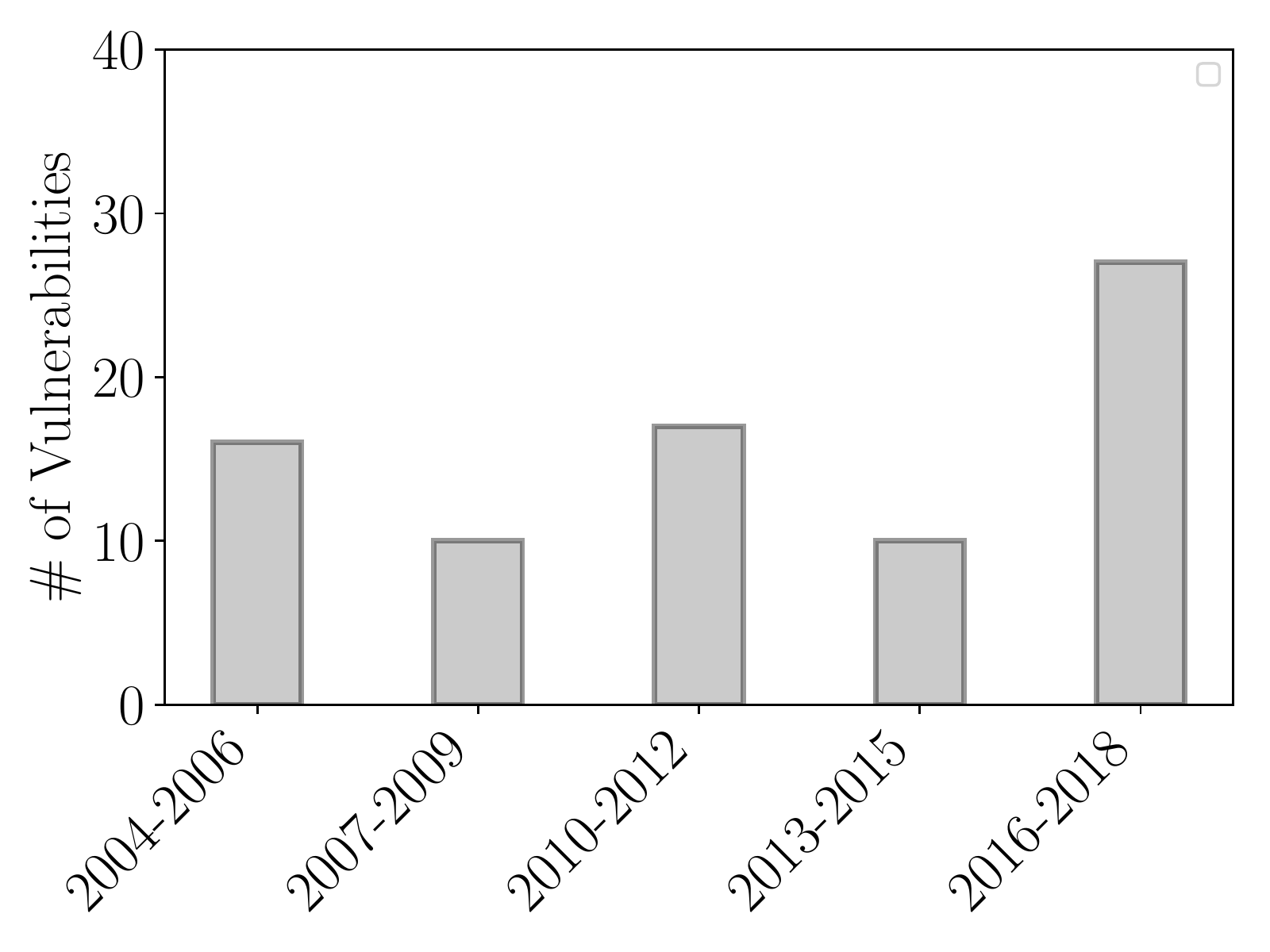}
\vspace{-0.2in}
\caption{Real-world vulnerabilities of our dataset over time.}
\label{fig:of-vulnerability-over-time}
\vspace{-0.03in}
\end{minipage}
\end{figure}

\subsection{Methodology}
\label{sec:methodology}

\begin{figure}[htbp]
	\centering
	\includegraphics[width=0.48\textwidth]{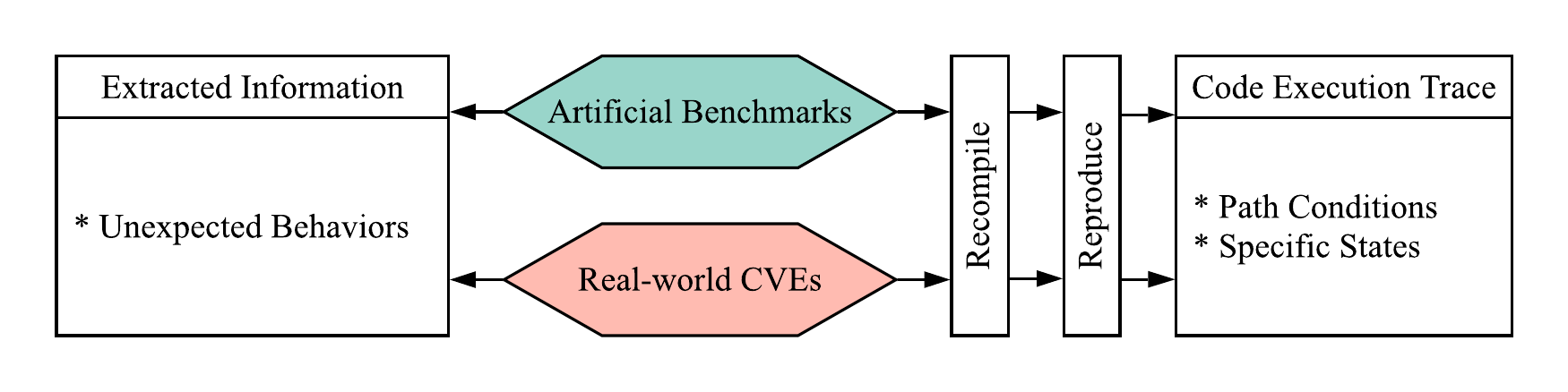}
	\caption{Workflow of the analyses of benchmarks and real-world vulnerabilities.}
	\label{fig:workflow}   
\end{figure}

We use different tools to collect the necessary data for our analysis.
The overall workflow for our data collection is shown in Figure~\ref{fig:workflow}.
As described in the workflow, we need to collect unexpected behaviors and code execution trace.

For the unexpected behaviors, we can obtain the types from the short description about the vulnerability and a list of external references on CVE website~\cite{CVEWeb}, which maintains a list of known vulnerabilities that have obtained a CVE ID.

\begin{figure*}[!htbp]
\centering
\begin{minipage}{0.47\linewidth}
 \includegraphics[width=\linewidth]{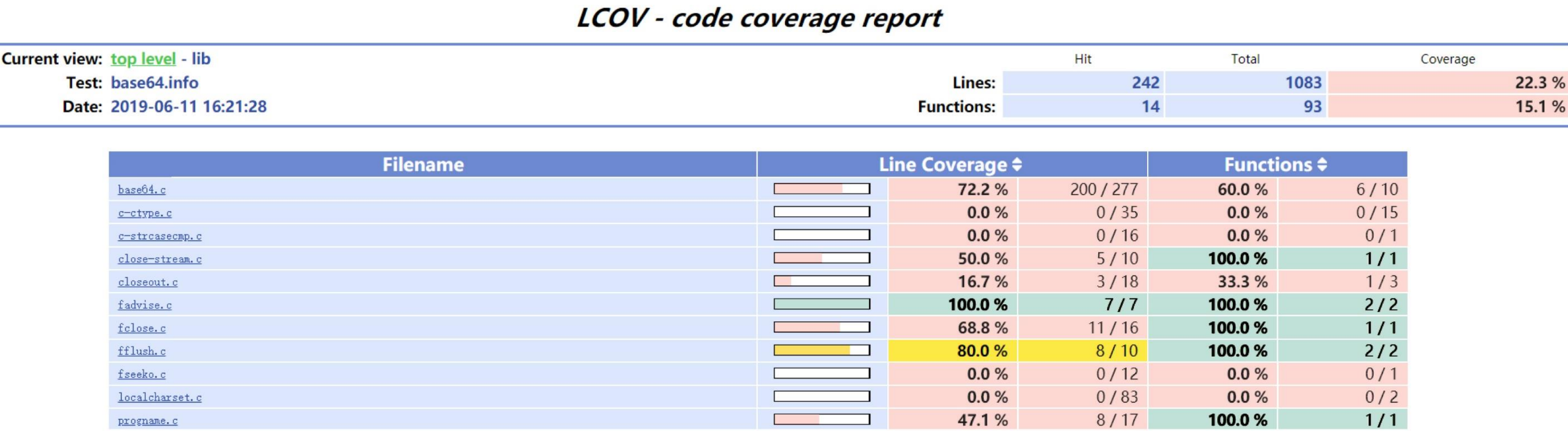}
\vspace{-0.2in}
\caption{An example of code coverage report generated by {\tt gcov} in {\tt base64} of LAVA-M.}
\label{fig:report}
\vspace{-0.03in}
\end{minipage}
\hfill
\begin{minipage}{0.45\linewidth}
\includegraphics[width=\linewidth]{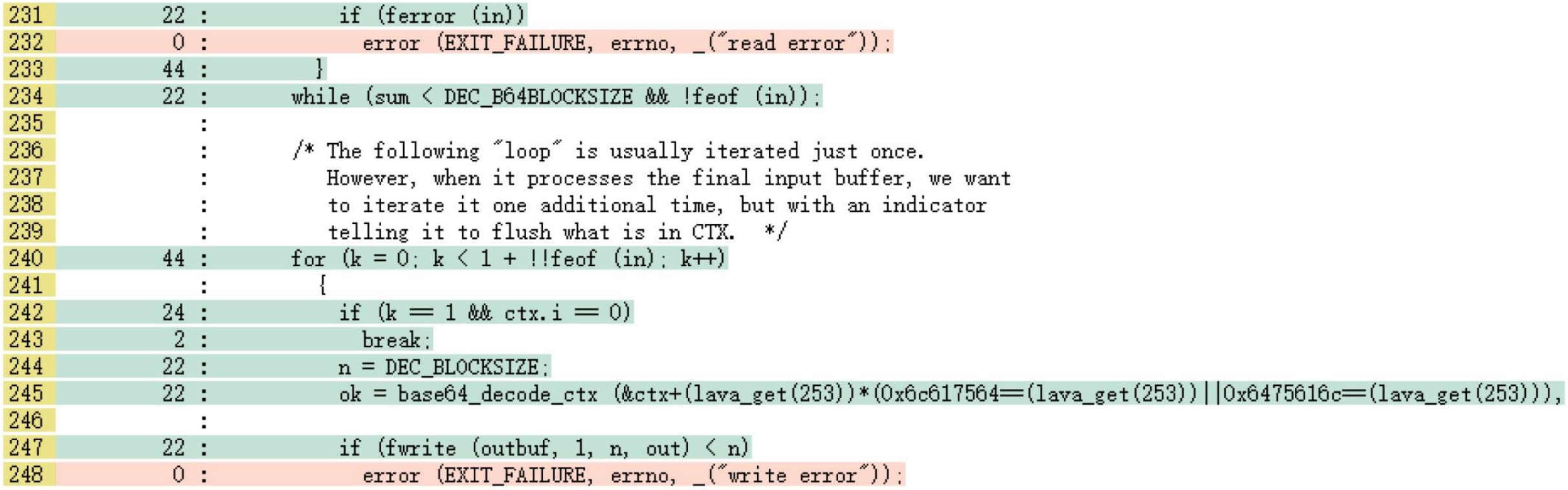}
\vspace{-0.2in}
\caption{An example of program execution trace in {\tt base64} of LAVA-M.}
\label{fig:code-coverage-example}
\vspace{-0.03in}
\end{minipage}
\end{figure*}

To take a deeper look into how the vulnerabilities occur and obtain the conditions for reaching the buggy locations and triggering the specific states, we need to collect the \emph{execution trace} of the vulnerability.
For this purpose, we need to \emph{recompile} the programs and \emph{reproduce} the vulnerabilities for their execution trace.
For all benchmarks, we first recompile these programs with {\tt gcov} flag --- a tool used in conjunction with GCC, to get the code coverage in programs.
Then we run the programs with the vulnerability inputs to get the \emph{code coverage report} and a typical code coverage report is shown in Figure~\ref{fig:report}. 
From the report, we can obtain the file name, the total and hit code lines and functions number, and the line coverage and function coverage of each program.
With the report, we can further extract the code execution traces of the vulnerabilities.
An example of the \emph{extracted execution trace} is shown in Figure~\ref{fig:code-coverage-example}.
In the code execution trace, the lines of code which have been executed are marked pink, while the unexecuted lines are marked green.
For the 80 CVEs, we collect all the vulnerability reports for each CVE and \emph{set up the operating system and environment}.
Because the vulnerabilities are different for each case and cannot be automated, it took us a lot of manpower(e.g., triggering and analyzing the crash step by step) to reproduce and analyze the vulnerabilities.
We manually analyze the \emph{path conditions} along the trace and the \emph{specific states} to fulfil for triggering the bug.

\section{Analysis results}
\label{sec:analysis_results}

We will give our analytic results about the comparisons between artificial benchmarks and real-world ones, and answer to the three research questions.

\subsection{(Q1) How do artificial vulnerabilities differ from real-world vulnerabilities? Can artificial vulnerabilities sufficiently mirror 
the reality?}
\label{sec:q1}

We depict the differences from two aspects: \textbf{requirements to trigger a vulnerability} and \textbf{unexpected behaviour types} of the buggy programs.

\subsubsection{Requirements to trigger a vulnerability}
As our model mentioned, a crash needs two requirements which are \emph{reaching the buggy location} and \emph{triggering specific states} along the execution trace.

\paragraph{Reach the buggy location}

Along the program execution, the input must meet a certain number of path conditions in terms of control flow and data flow to reach the buggy location.
Some of the paths are protected by magic bytes comparisons, and can only be reached when the detection tools solve
the corresponding constraints.
Also, the input may be go through some data transformations(e.g., arithmetic and hash operations).
So we first analyze the types and proportions of magic numbers in the constraints.
Then we analyze the proportion of data transformation in the trace between artificial benchmarks and real-world vulnerabilities.

\ding{182}The types and proportions of magic numbers.

Magic numbers refer to the bytes in the test input which are uses in comparison instructions.
The types of magic numbers influence the difficulty of solving the comparisons. 
We summarize the types of magic numbers and their
proportions in Table~\ref{tab:Magic-number-type-of-datasets}.
For our dataset, the comparisons in LAVA-M and Rode0day cover only two types
and are almost 32-bit. In contrast, CGC and 
real-world programs have various types of magic numbers.
When we further
study the proportion of each type, we find that
though LAVA-M and Rode0Day put nearly 90\% of their efforts on
the type of {\tt int32}, it only accounts for 17\% in the 
real world. 

To summarize, since the distribution and the types of 
magic numbers in LAVA-M and Rode0day are far away from the
real-world vulnerabilities, they may not 
evaluate the ability to solve constraints of 
the vulnerability detection tools as fairly as the real world.
As CGC is written by specialists, their magic numbers can
be designed more freely and seem more realistic.

\begin{table}[h]
\centering
\caption{Magic number type of datasets.}
\label{tab:Magic-number-type-of-datasets}
\begin{tabular}{c|c|c|c|c}
\bottomrule[1.3pt]
\textbf{\begin{tabular}[c]{@{}c@{}}Magic Number \\ Type\end{tabular}} & \textbf{LAVA-M} & \textbf{Rode0day} & \textbf{CGC} & \textbf{Real World} \\ \hline
\textbf{char}                                                         & 8\%             & 10\%              & 40\%         & 51\%                \\ \hline
\textbf{string}                                                       & -                & 3\%               & 29\%         & 26\%                \\ \hline
\textbf{float}                                                        & -                & -                  & 5\%          & -                    \\ \hline
\textbf{unsigned int16}                                               & -                & -                  & 0.5\%        & 1\%                 \\ \hline
\textbf{unsigned int32}                                               & -                & -                  & 3\%          & 4\%                 \\ \hline
\textbf{unsigned int64}                                               & -                & -                  & 0.5\%        & -                    \\ \hline
\textbf{int32}                                                        & 92\%              & 87\%              & 21.5\%       & 17\%                \\ \hline
\textbf{long int}                                                     & -                & -                  & 0.5\%        & 1\%                 \\ 
\bottomrule[1.3pt]
\end{tabular}
\end{table}




\ding{183}The proportion of data transformation in the trace.

Data transformation means the arithmetic or hash operations that the input may have gone through.
We analyze the input's transformation along the program execution between artificial benchmarks and real-world vulnerabilities.  
At the points where LAVA-M and Rode0day insert bugs, the values of magic numbers are often the direct copies of inputs without any data transformation. 
Our statistics of the real world and CGC in Table~\ref{tab:data-transformation-of-datasets} show that 61.8\% of inputs in real-world programs have data transformations, including but not limited to type conversion, casting the integer to the character or doing inversely and bit-wise operation in real-world. 
As to CGC, only 40\% of them have data transformations along the trace, where type conversion is the majority.

\begin{table}[htbp]
\centering
\setlength{\tabcolsep}{5mm}{
\caption{The proportion of data transformation of datasets.}
\label{tab:data-transformation-of-datasets}
\begin{tabular}{c|c|c}
\toprule[1.3pt]
\multicolumn{1}{c|}{\textbf{Dataset}} & \multicolumn{1}{c|}{\textbf{Transformation}} & \multicolumn{1}{c}{\textbf{Non-Transformation}} \\ \hline
LAVA-M                                  & 0\%                                          & 100\%                                           \\ \hline
Rode0day                              & 8\%                                          & 92\%                                           \\ \hline
CGC                                   & 40.0\%                                       & 60.0\%                                          \\ \hline
Real World                            & 61.8\%                                       & 38.2\%                                          \\ 
\bottomrule[1.3pt]
\end{tabular}}
\end{table}



\paragraph{Trigger specific program states}

As have mentioned before, since LAVA-M and Rode0day artificially construct the relation between DUA and ATP, the execution trace to trigger the vulnerability must fulfill the series of path conditions designated by them, including the magic number guarded at the last path comparison.
As to LAVA-M, once the magic number of the last if-condition is solved out, the program will crash immediately without any specific state to be triggered along the trace.
Though Rode0day has improved LAVA by using multiple DUAs in the last conditions to complicate the comparisons, its vulnerability triggering mechanism still excludes any specific state.

The vulnerabilities in the real world and CGC reveal the opposite feature at the time of crash.
To make a memory corruption vulnerability, it must go through some specific states and trigger that.
For example, as for Buffer-Overflow vulnerability, the program should read or write a longer size of data to the buffer allocated, that is, if it triggers the specific states, the program will crash successfully.
As for use-after-free vulnerability, before the pointer used, it must be freed first, which triggers a specific state.

\begin{figure}[!ht]
\begin{lstlisting}[language=c,
xleftmargin=.05\columnwidth, xrightmargin=.01\columnwidth,
% xleftmargin=.08\columnwidth, xrightmargin=.03\columnwidth,
% linewidth=.95\columnwidth, 
linebackgroundcolor={%
\ifnum\value{lstnumber}>4\ifnum\value{lstnumber}<6\mtcolor\fi\fi
\ifnum\value{lstnumber}>11\ifnum\value{lstnumber}<14\mtcolor\fi\fi
}]
void receive_input(FILE *f){
    ...
    char buf[1024];
    fread(buf, 1024, 1, f);
    lava_set(*(int *)buf);
    ...
}
...
void foo(char *bar){
    // BUG:

    printf("%s\n", bar + lava_get() * 
    (lava_get() == 0x6176616c));
}
\end{lstlisting}
\caption{A bug inserted by LAVA-M.}
\label{lis:lava-bug}
\end{figure}

\begin{figure}[ht]
\begin{lstlisting}[language=c,
xleftmargin=.05\columnwidth, xrightmargin=.01\columnwidth,
% xleftmargin=.08\columnwidth, xrightmargin=.03\columnwidth,
% linewidth=.95\columnwidth, 
linebackgroundcolor={%
\ifnum\value{lstnumber}>4\ifnum\value{lstnumber}<6\mtcolor\fi\fi
\ifnum\value{lstnumber}>8\ifnum\value{lstnumber}<10\mtcolor\fi\fi
}]
void ReadFace(fbuf){
    ...
    t = s = fbuf;
    if(...){
        char buffer[128];
        sscanf(s, "%s %s", buffer);}
        ...
    c = (int)* (s++);
    if((c >= '0') && (c <= '9'))
    ...
\end{lstlisting}
\caption{A buffer overflow vulnerability in CVE-2009-2286.}
\label{lis:cve-bug}
\end{figure}

In the following, we present a detailed \textit{case study} to illustrate the difference in triggering specific states between benchmarks and real-world vulnerabilities.

Figure~\ref{lis:lava-bug} shows an example bug inserted by LAVA-M.
The function {\tt lava\_get} retrieves the value last stored by a call to {\tt lava\_set}.
Once the magic number {\tt 0x6176616c} is solved out, the program will crash even without any specific state to be triggered as the highlighted code line 12 and 13.

As for real world, taking CVE-2009-2286 in Figure~\ref{lis:cve-bug} as an example, we can observe that
when the input solely reaches the last condition 
{\tt if((c>='0')\&\&(c<='9'))} as the highlighted code line 9,
the buffer overflow vulnerability in {\tt ReadFace} may \emph{not} be triggered. 
Unless triggering the specific state in the highlighted code line 5, that is, the size of input is longer than 128 that the buffer allocated, the program will only crash successfully.

From the case study, we can draw the conclusion that the bugs artificially inserted by LAVA-M and Rode0day,
their triggering conditions are deliberately designed and do not include specific states to trigger, which distinct from the true circumstances under the real world. 
Such design may not well evaluate the ability of 
\emph{program state coverage} exploration for vulnerability detection tools.

\subsubsection{Unexpected behaviour types} 
Unexpected behaviour types, also called vulnerability type, can evaluate the sensitivity of vulnerability detection tools.
If the tools are less effective, they can not detect diverse vulnerability types.

For the vulnerabilities of real world, we look up to the 5 most common memory corruption vulnerabilities in real world~\cite{CWEWeb}. 
As Table~\ref{tab:CGC-Realworld-Bug-Type} shows, the top 3 bugs type of CGC are Out-of-Bounds Write, Out-of-Bounds Read, and Null-Pointer-Dereference, which account for 37.8\%, 22.7\% and 20.2\% of the total bugs we collected respectively. 
Compared to the real world, however, the same bug types only account for 3.4\%, 8.6\% and 5.0\% respectively, which indicates the great difference of the bug type distribution between CGC and real world. 
Though CGC has many types of bugs, the percents are much less than the real-world circumstances.
A probable reason why CGC has more Out-of-Bounds Writing and Reading bugs is that they are easier to create.
As mentioned above, LAVA and Rode0day can inject \emph{only} buffer overflow bugs into programs. 

The benchmarks that lack diversity in inserted bugs may not well
evaluate the \emph{sensitivity} of vulnerability detection tools, since certain
bugs excluded from the benchmarks may not be detected by some
less effective tools.

\begin{table}[hbp]
\scriptsize
\centering
\caption{The 5 most common software weaknesses of memory corruption vulnerabilities in real world~\cite{CWEWeb}, and the number of them in CGC.}
\label{tab:CGC-Realworld-Bug-Type}
\begin{tabular}{c|c|c|c}
\toprule[1.3pt]
   & \textbf{\begin{tabular}[c]{@{}c@{}}Weakness \\ Summary\end{tabular}}                                          & \textbf{\begin{tabular}[c]{@{}c@{}}Real-World \\ Num  (Percentage)\end{tabular}} & \textbf{\begin{tabular}[c]{@{}c@{}}CGC \\ Num  (Percentage)\end{tabular}} \\ \hline
1              & Buffer-Overflow          & 12293 (76.3\%)                                                                      & 14 (11.7\%)                                                                  \\ \hline
2              & Out-of-Bounds Read                                                 & 1390 (8.6\%)                                                                        & 27 (22.7\%)                                                                  \\ \hline
3              & Use-After-Free                                                     & 1079 (6.7\%)                                                                        & 9 (7.6\%)                                                                    \\ \hline
4              & \begin{tabular}[c]{@{}c@{}}Null-Pointer\\ -Dereference\end{tabular} & 805 (5.0\%)                                                                         & 24 (20.2\%)                                                                  \\ \hline
5              & Out-of-Bounds Write                                                & 548 (3.4\%)                                                                         & 45 (37.8\%)                                                                  \\ 
\bottomrule[1.3pt]
\end{tabular}
\end{table}

\begin{tcolorbox}[size=title, opacityfill=0.15]
Though some of the features share similarities, there still exist significant differences between artificial benchmarks and reality, which may influence the evaluation of the vulnerability detection tools.
\end{tcolorbox}

\subsection{(Q2) What can we find by modifying the artificial benchmarks according to what we have observed from Q1?}
\label{sec:q2}

The above analytic results indicate that the vulnerability
features in the artificial benchmarks differ greatly from
the real world. Thus, we further carry on a series of experiments 
to evaluate how these differences will affect the utilities of
vulnerability detection tools. 
We modify the artificial benchmarks in accordance with
the features we summarized for the real-world vulnerabilities,
and test them with the vulnerability detection tool AFL~\cite{AFL}.
Our experiments are only conducted on LAVA-M because
Rode0day is evolved from LAVA with a similarly unsatisfactory
manifestation in our model, 
while CGC is written by specialists and its features are close
to the reality in most metrics.

First, we modify the program in LAVA-M to make it more realistic from the following aspects:
\ding{182}We extend the magic number types to {\tt char}, {\tt string} and {\tt int32} (we omit the types that account for less 
than 5\% in Table~\ref{tab:Magic-number-type-of-datasets}) by modifying \emph{if conditions}, with a similar proportion to real-world cases.
\ding{183}We add extra \emph{data transformation} to the DUAs before they are compared to the magic number.
\ding{184}We combine \emph{both} the above modifications.

Considering the high similarity between the LAVA-inserted bugs in 
different programs, we only modify {\tt base64} in LAVA-M.
After applying these modifications, we obtain 3 different versions 
of {\tt base64}, and run AFL on them respectively, as well as the 
original {\tt base64} in LAVA-M.
Each of the experiments is repeated 5 times to mitigate the randomness of fuzzing and lasts for 5 hours to align with~\cite{dolan2016lava,li2017steelix,rawat2017vuzzer}.

Lastly, we collect the unique crashes detected by AFL in the experiments to evaluate the effects of the modifications.

\begin{figure}[h]
	\centering
	\includegraphics[width=0.35\textwidth]{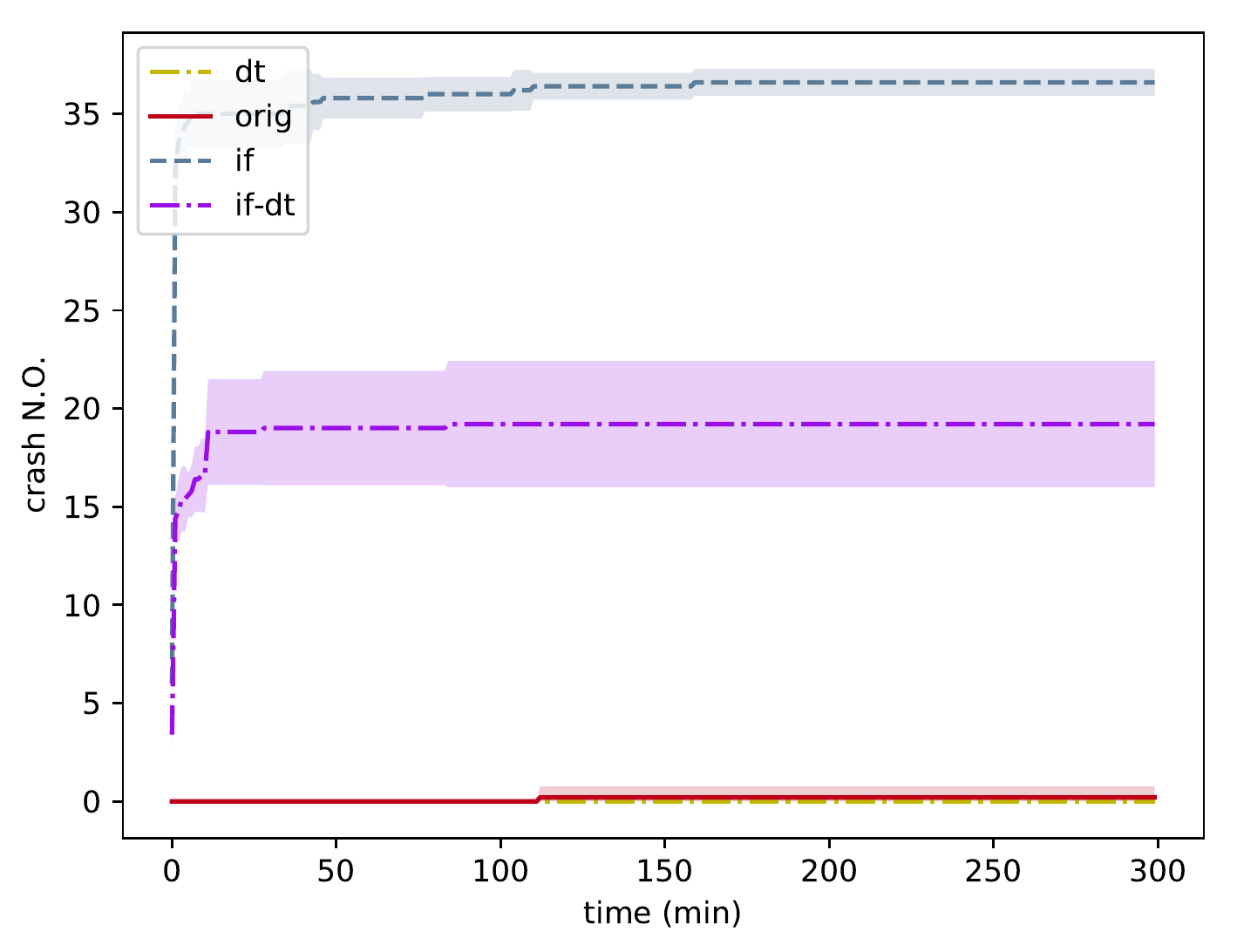}
	\caption{Crashes detected by AFL in 5 hours. The shaded area is the 95\% confidence interval. ``if" means the if conditions are modified; ``dt" means data transformation is added; ``if-dt" means both are modified; ``orig" means original {\tt base64}.}
	\label{fig:crashes_base64}   
\end{figure}

Figure~\ref{fig:crashes_base64} shows the average number of unique crashes detected by AFL during the experiments.
We can see that modifying the \textit{if conditions} makes it much easier for AFL to detect the bugs.
This is because almost half of the comparisons in real-world cases are \texttt{char} comparisons, which are easier for AFL to penetrate.
Adding \textit{data transformations} makes it harder for AFL to detect bugs. 
In fact, AFL detects no bug in this version of {\tt base64}.
However, the difference between this version and the original version is minor as the 95\% confidence intervals overlap and the p-value of Mann-Whitney U test is 0.212, above 0.05.
Because the bugs are already hard for AFL to detect, making it harder does not cause big differences.
The version applying both \textit{if conditions} modification and \textit{data transformation} addition is closest to real-world cases.
The number of unique crashes detected is between the other 2 modified versions as expected.
However, the results on this realistic version differ very much from the results on the original version as the confidences intervals are discrete and the p-value is 0.005, below 0.05.
The results of the experiments show that the artificial bugs cannot mirror the reality well because they may not fairly reflect the properties of the vulnerability detection tool.

\begin{tcolorbox}[size=title, opacityfill=0.15]
Through the above analysis, we can see the benchmarks and the real-world vulnerabilities have different manifestations on our model and can not evaluate vulnerability detection tools well in the aspects of program state coverage and sensitivity. 
\end{tcolorbox}

\subsection{(Q3) What improvements can we make towards more realistic artificial vulnerabilities benchmarks?}
\label{sec:q3}

In this section, we discuss how to make the artificial benchmark more realistic based on the program perspective and vulnerability.
We can:
\ding{182} \emph{add different types of magic numbers}.
This is because we find that the magic bytes which are uses in comparison instructions are diverse and they are a range but not a definite number.
Adding a hard-to-match magic number along the path can make it harder to vulnerability techniques for detection but cannot represent the path conditions in real-world programs.
\ding{183} \emph{add some data transformations} along the execution trace of vulnerabilities.
We found that the input data goes through data transformations on real-world vulnerabilities.
This is complementary to distributing the path conditions, which also helps to test out the program coverage capability of vulnerability detection tools.
\ding{184} \emph{set proper special states} as requirements for triggering the artificial vulnerabilities.
We found that only reaching the buggy location is often not enough to trigger the vulnerabilities in real-world cases.
An artificial vulnerability should have a certain program state as its prerequisite to better represent the real-world case.
This can help to test out the program state exploration capability of the vulnerability detection tools and is complementary to the two previous suggestions which are related to making the buggy location hard to reach.
\ding{185} \emph{add different types of vulnerabilities}.
Different types of vulnerabilities often require different approaches for effective detection.
This can help to test the versatility and sensitivity of the detection tool.

\begin{tcolorbox}[size=title, opacityfill=0.15]
With the model and the analysis of different artificial benchmarks as well as programs, we provide suggestions from the vulnerability perspectives for building a more realistic benchmark.

\end{tcolorbox}

\section{Threats to Validity}

There are some potential threats to the validity of this study.
Foremost is simply the representative of all memory corruption vulnerabilities in real world.
We believe, however, that we have chosen a representative set of real-world vulnerabilities and we will argue from four aspects.

\textbf{Vulnerability Type.}
The first potential threat is the limited scope of the study in terms of the vulnerability types. 
In this paper, we focus on the memory corruption vulnerabilities
as they have been ranked among the most dangerous software errors~\cite{CWEdanger} due to their high severity and real-world impact~\cite{ms_mc}.
There are more than 10,000 memory corruption vulnerabilities listed on the CVE website. We crawled the pages of the current 95K+ entries (2001-2017) and analyzed their severity scores (CVSS). 
Our result shows that the average CVSS score for memory corruption vulnerabilities is 7.6, which is clearly higher than the overall average (6.2), confirming their severity.


\textbf{Justifications on the Dataset Size.}
The second potential threat to validity is that the size of our dataset, specifically the number of real-world vulnerabilities.
We select 80 CVEs as real-world vulnerabilities and argue that the size of our dataset is reasonably large.
For example, recent work that uses vulnerabilities in the CVE list to evaluate their vulnerability detection systems are limited to less than 10 vulnerabilities~\cite{carlini2015control, hu2016data, castro2006securing, akritidis2008preventing}, due to the manual efforts needed to build ground truth data.
So considering both the number and manual efforts, we believe the size of dataset is enough.

\textbf{Others.}
Among the three most-used artificial benchmarks, although CGC is the closest to the real-world vulnerabilities, it also has some shortcomings.
For example, we obtain cyclomatic complexity (CCN) of the programs in artificial benchmarks and real world. 
If the source code contained more control-flow statements (conditionals or decision points), the complexity would be more.
The results show that the average CCN values of Rode0day (9.1), LAVA-M (8.6), and real-world programs (8.1) are higher than the value of CGC (5.2).
The median CCN values of LAVA-M and Rode0day are similar to real-world ones because they exactly insert vulnerabilities to real-world programs.
The result indicates the logic inside these specialists' crafts (CGC) is simpler than the real-world cases.
We need to do much more exploratory analysis to determine how it influence the vulnerabilities hidden in the small-sized benchmarks to be detected by vulnerability detection tools compared with the real-world ones.

\section{Related work}
\label{sec:related}

Since this work is the first one to systematically study the artificial
and real-world vulnerabilities and explain how benchmarks evaluate 
the vulnerability detection tools, we focus on the related work from three categories: the artificial benchmarks, studies about real-world vulnerabilities and the vulnerability detection tools.

\textbf{Artificial Benchmarks.}
There are some artificial benchmarks to evaluate vulnerability detection tools~\cite{dolan2016lava, pewny2016evilcoder, Rode0dayWeb, CGCWeb, hu2018chaff, roy2018bug}.
Besides the three most-used benchmarks we study,
Evilcoder~\cite{pewny2016evilcoder} automatically find potentially vulnerable source code locations and transform benign code into vulnerable code.  
Bug Synthesis~\cite{roy2018bug} uses constraint-based program synthesis to automatically inject bugs.
Unlike the above works which focus on building new benchmarks,
we concentrate on systematically studying the differences between the three most-used artificial benchmarks and real-world vulnerabilities.

\textbf{Studies of Real-world Vulnerabilities}
Several works have been proposed to study the properties of real-world vulnerabilities~\cite{walden2014, fabian2014, leopard2018}.
Their main purposes are different from ours.
They focus on detecting real-world vulnerabilities with the metrics while we focus on comparing real-world vulnerabilities with artificial bugs.

\textbf{Vulnerability Detection Tools.}
The major purpose of benchmarks is to evaluate vulnerability detection tools.
Rise of the HaCRS~\cite{shoshitaishvili2017rise} and  Driller~\cite{stephens2016driller} use CGC as the benchmark. 
VUzzer~\cite{rawat2017vuzzer}, Steelix~\cite{li2017steelix} and QSYM~\cite{yun2018qsym} use LAVA-M as the ground-truth. 
But few published vulnerability detection tools use Rode0day for evaluation because it is a very recent work. 
The aforementioned tools also use real-world programs as ground truth, like {\tt nm}, {\tt objdump} and {\tt gif2png}, etc.

\section{Future Work}

Our future work primarily focuses on implementing a new benchmark. 
For example, our findings suggest that adding data transformations along the trace and making a uniform difficulty for reaching path conditions are more realistic for creating bugs artificially. 
Broadening the diversity of types of bugs and injecting exploitable bugs explicitly are also future directions for our work. 

Moreover, we will explore the current state of vulnerability discovery and software security based on our analysis results, like what properties make a given bug difficult to find and what types of bugs are currently missed by existing vulnerability finding tools. 
This can inform us of how to improve the state-of-the-art in the field of vulnerability discovery.  

\section{Conclusion}
\label{sec:conclusion}
This work presents an in-depth empirical study on artificial vulnerability 
benchmarks, with the goal of understanding how close these
benchmarks represent the reality. To maintain the generality and 
objectiveness, our study covers three most-used artificial 
benchmarks of 2669 bugs and a diverse group of 80 real-world vulnerabilities. Our study follows 
a general model of vulnerabilities and systematically compares 
the artificial vulnerabilities against real ones. In short, 
our study centers around three questions and finds
that while artificial benchmarks attempt to approach the reality, 
they still significantly differ from the real world in several aspects. Last but 
not least, following our study, we modify LAVA-M and propose a set of
strategies towards making artificial benchmarks more realistic.

\bibliography{references}
\bibliographystyle{IEEEtranN}

\end{document}